\documentclass[aps,prl,twocolumn,superscriptaddress,floatfix,citeautoscript]{revtex4}
\usepackage{dcolumn}
\usepackage{bm}
\usepackage{amsmath,amssymb}
\usepackage{times}
\usepackage{float}
\usepackage{graphicx, epstopdf}

\begin{document}

\title{Supplemental Material for ``High Contrast X-ray Speckle from Atomic-Scale Order in Liquids and Glasses''}

\author{S. O. Hruszkewycz}
\affiliation{Materials Science Division, Argonne National Laboratory, Argonne, Illinois 60439 USA}
\author{M. Sutton}
\affiliation{Department of Physics, McGill University, Montreal, H3A2T8 Canada}
\author{P. H. Fuoss}
\affiliation{Materials Science Division, Argonne National Laboratory, Argonne, Illinois 60439 USA}
\author{B. Adams}
\affiliation{X-ray Science Division, Argonne National Laboratory, Argonne, Illinois 60439 USA}
\author{S. Rosenkranz}
\affiliation{Materials Science Division, Argonne National Laboratory, Argonne, Illinois 60439 USA}
\author{K. F. Ludwig~Jr.}
\affiliation{Department of Physics, Boston University, Boston, Massachusetts, 02215 USA}
\author{W. Roseker}
\affiliation{HASYLAB, Deutsches Elektronen-Synchrotron, Hamburg, Germany}
\author{D. Fritz}
\affiliation{Linac Coherent Light Source, SLAC National Accelerator Laboratory, Menlo Park, California 94025 USA}
\author{M. Cammarata}
\affiliation{Linac Coherent Light Source, SLAC National Accelerator Laboratory, Menlo Park, California 94025 USA}
\author{D. Zhu}
\affiliation{Linac Coherent Light Source, SLAC National Accelerator Laboratory, Menlo Park, California 94025 USA}
\author{S. H. Lee}
\altaffiliation{Present address: Physical Metrology Division, Korea Research Institute of Standards and Science, Daejeon 305-340 S. Korea.}
\affiliation{HASYLAB, Deutsches Elektronen-Synchrotron, Hamburg, Germany}
\affiliation{Linac Coherent Light Source, SLAC National Accelerator Laboratory, Menlo Park, California 94025 USA}
\author{H. Lemke}
\affiliation{Linac Coherent Light Source, SLAC National Accelerator Laboratory, Menlo Park, California 94025 USA}
\author{C. Gutt}
\affiliation{HASYLAB, Deutsches Elektronen-Synchrotron, Hamburg, Germany}
\author{A. Robert}
\affiliation{Linac Coherent Light Source, SLAC National Accelerator Laboratory, Menlo Park, California 94025 USA}
\author{G. Gr\"{u}bel}
\affiliation{HASYLAB, Deutsches Elektronen-Synchrotron, Hamburg, Germany}
\author{G. B. Stephenson}
\affiliation{Materials Science Division, Argonne National Laboratory, Argonne, Illinois 60439 USA}
\affiliation{Advanced Photon Source, Argonne National Laboratory, Argonne, Illinois 60439 USA}

\date{\today}

\begin{abstract}  
This supplemental material gives additional detail on Experimental Methods and Hard X-ray FEL Source Characteristics,
Calculation of Maximum Speckle Contrast, Extracting Contrast of Weak Speckle Patterns,
Estimated Temperature Increase from X-ray Absorption, 
Split-Pulse XPCS Feasibility, and Sample Disturbance During Single Pulses.
\end{abstract}

\maketitle

\section{Experimental Methods and Hard X-ray FEL Source Characteristics}

All data presented are from experiment L264, 
which was one of the first hard x-ray experiments at LCLS,
and used the first available hard x-ray instrument (XPP).
Improvements can thus be expected in future experiments
due to accumulating experience in operating the 
LCLS to produce hard x-rays,
and use of the new XCS instrument
designed for XPCS measurements.

A double-crystal Si (111) monochromator was used to select a photon energy of 7.99 keV.
Incident x-ray pulse energies (photons per pulse)
were determined from the IPM2 monitor
downstream of the monochromator but upstream of the attenuators,
calibrated against an x-ray pulse power meter placed at the sample position. 
When attenuators were used, the calculated attenuation factors were applied
to determine the pulse energy $\langle I_0 \rangle$ incident on the sample.
Beryllium compound refractive lenses with a focal length of 4 m were used 
to focus the beam to a $1.6 \times 1.7~\mu$m H$\times$V FWHM spot at the sample position.
The focal spot size was measured by scanning slit edges 
across the attenuated beam.
The stability of the focal spot position was $0.25 \times 0.37~\mu$m H$\times$V FWHM
based on an analysis of the downstream signal fluctuations observed
with the slit edges cutting half of the beam.

The sample was mounted in a helium-filled chamber.
For the liquid Ga, 
a 2.8-mm diameter hemispherical droplet was mounted on a heated substrate,
with the beam impinging upstream of and below its apex to produce
an approximately symmetric reflection scattering geometry.
For the Ni$_2$Pd$_2$P glass,
a sample with a polished surface inclined at 15 degrees
was used to give an approximately symmetric reflection scattering geometry.
For the B$_2$O$_3$ glass, 
a fiber of diameter $18~\mu$m was suspended across an aperture
and positioned in transmission geometry. 
On each sample, several locations were investigated,
and data from different locations were typically combined in the analysis.
For the unattenuated single-pulse Ni$_2$Pd$_2$P data,
where intense pulses would permanently damage the surface (see the final section below),
the sample was displaced by 40 $\mu$m every 1 or 5 pulses so as to scatter from a new sample region. 
A long-focal-length microscope fitted with a video camera
allowed real-time observation of sample damage by the focused x-ray beam.
Scattering experiments were carried out with a CCD detector 
having a direct-x-ray-detection deep-depletion silicon sensor 
with $1340 \times 1300$ pixels $20~\mu$m square, 
mounted so that it could be positioned in a vertical scattering plane
at various distances $L$ from the sample. 
Helium flight paths were used for longer detector distances.

The free electron laser parameters used were 250 pC bunch charge, 
a 70 fs electron beam integral pulse width, and a 60 Hz repetition rate.
The machine monitors typically reported $2 \pm 1$ mJ nominal pulse energy
integrated over the full x-ray photon energy spectrum.
A taper was introduced into the final segments of the undulator which was observed to optimize the
monochromatic x-ray pulse energy.
To accommodate the $\sim 2$ second readout time of the CCD
without the availability of a detector shutter,
we operated the LCLS in ``burst mode'',
where a burst containing a specified number of electron pulses
(at a 60 Hz repetition rate)
could be delivered to the undulator,
with timing coordinated so that the CCD image could be recorded between bursts.
Including the overhead associated with handshaking between various components
of the x-ray source and data acquisition system,
the maximum rate at which CCD patterns from a sequence of bursts could be recorded (even with single pulses)
was one per six seconds.

\begin{figure}
\includegraphics[width=3.0in]{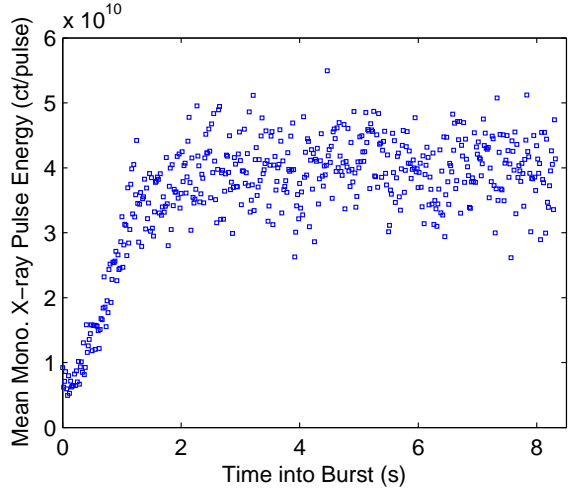}
\caption{(Supplemental) Mean monochromatic x-ray pulse energy (photons per pulse) averaged over 50 bursts,
as a function of time into each 500-pulse burst.
The initial pulses in a burst were typically much weaker
than later pulses.}
\end{figure}

\begin{figure}
\includegraphics[width=3.0in]{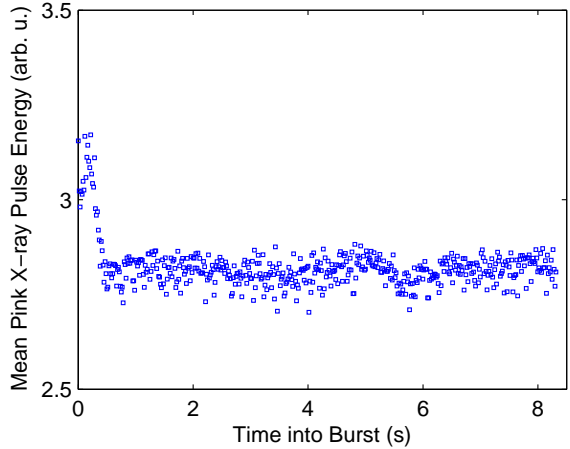}
\caption{(Supplemental) Mean pink (un-monochromated) x-ray pulse energy averaged over 50 bursts, 
as a function of time into each 500-pulse bust.  
In contrast to the monochromatic results,
the initial pink x-ray pulses in a burst have about 10\% more pulse energy.}
\end{figure}

\begin{figure}
\includegraphics[width=3.0in]{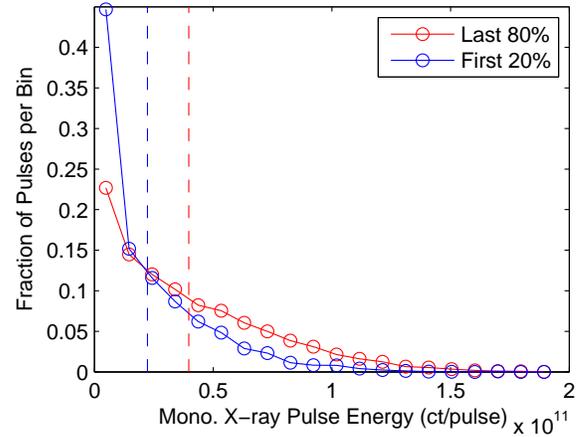}
\caption{(Supplemental) Histogram of monochromatic x-ray pulse energies
in last 80\% and first 20\% of burst,
for fifty 500-pulse bursts.
Both show distributions peaked at zero.
Maximum pulse energies were about 5 times larger than
the mean pulse energies (given by vertical dashed lines).}
\end{figure}

\begin{figure}
\includegraphics[width=3.0in]{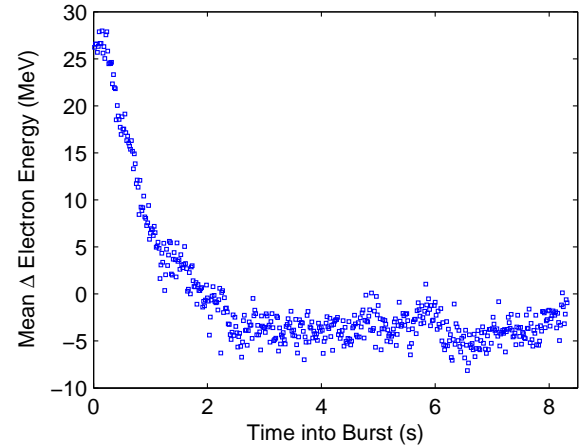}
\caption{(Supplemental) Mean deviation (in MeV) of nominal electron beam energy 
in undulator from its global average of 13,426 MeV,
as determined by the electron beam L3 monitor,
averaged over 50 bursts,
as a function of time into each 500-pulse burst.
The nominal electron energy of the initial pulses in a burst 
deviated by about 30 MeV from its eventual value.}
\end{figure}

\begin{figure}
\includegraphics[width=3.0in]{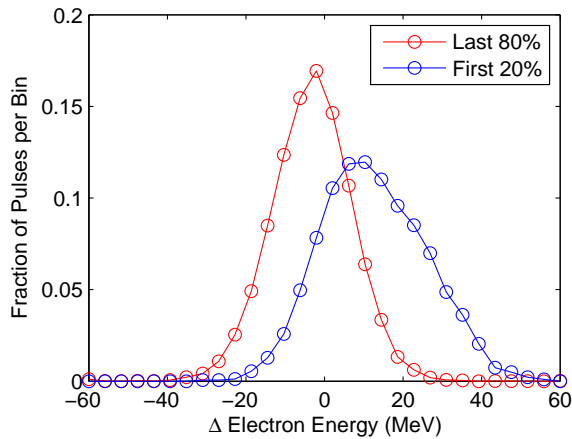}
\caption{(Supplemental) Histogram of electron beam energy deviation of each pulse 
in last 80\% and first 20\% of burst,
for fifty 500-pulse bursts.
The width of the former distribution is 22 MeV FWHM.}
\end{figure}

\begin{figure}
\includegraphics[width=3.0in]{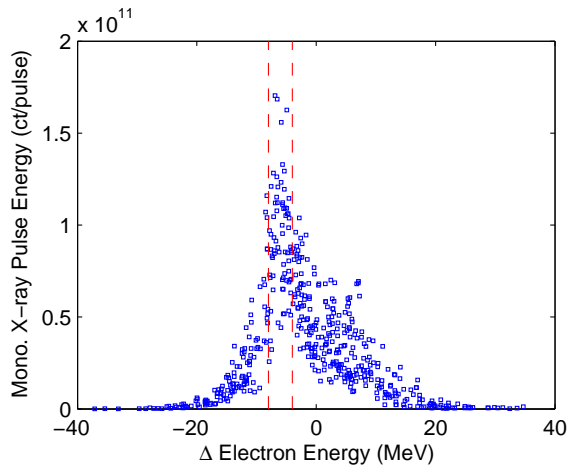}
\caption{(Supplemental) Monochromatic x-ray pulse energy plotted vs. electron beam energy deviation
for pulses in a single 500-pulse burst.
Vertical dashed lines show the electron energy range used to select the pulses
histogrammed in Fig. 8 (Suppl.).}
\end{figure}

\begin{figure}
\includegraphics[width=3.0in]{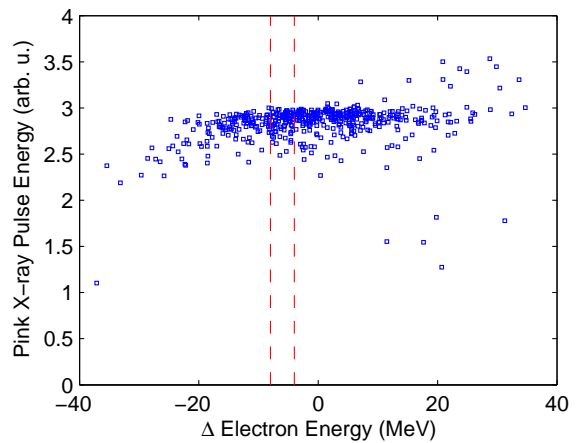}
\caption{(Supplemental) Pink (un-monochromated) x-ray pulse energy plotted vs. electron beam energy deviation 
for pulses in a single 500-pulse burst.
Vertical dashed lines show the same electron energy range as in Figure 6 (Suppl.). }
\end{figure}

Non-optimal machine tuning of burst mode during this experiment 
resulted in lower average monochromatic x-ray pulse energy
for bursts of fewer than about 100 pulses than for longer bursts. 
The average energy of the LCLS electron beam is stabilized by a feedback loop, 
and different signals are used depending on whether or not  x-rays are being produced during a burst.
During this experiment, these signals were not properly calibrated against each other, 
resulting in a transient change in the average electron beam energy at the beginning of bursts. 
While this problem was later diagnosed
and a method of calibrating the signals to eliminate the transient 
and obtain high x-ray pulse energy in short bursts 
was implemented for later experiments,
we describe the problem here to explain why our single-pulse results
have lower than optimal incident intensity.

Figure 1 (Supplemental) shows the mean monochromatic x-ray pulse energy
as a function of time during 500-pulse bursts, averaged over 50 bursts.
(Data shown in this section are from run 206,
and are typical of all runs in this experiment.)
On average, the initial pulses in a burst were significantly less intense than later pulses 
despite the fact that the un-monochromated (pink) x-ray pulse energies 
were initially higher than average, as shown in Fig. 2 (Suppl.). 
This indicates that the initial pulses had a different average photon energy than later pulses.
Figure 3 (Suppl.) shows the distribution of the monochromatic x-ray  pulse energies,
for both the initial 100 and final 400 pulses of 500-pulse bursts.
Both show distributions peaked at zero;
the mean shown by the vertical dashed line 
was higher for the later pulses.

The transient in the monochromatic x-ray pulse energy at the beginning of a burst
is related to the energy of the electron beam in the undulator.
Figure 4 (Suppl.) shows the nominal electron beam energy
as a function of time during the bursts,
averaged as in Fig. 1 (Suppl.),
while Fig. 5 (Suppl.) shows the distribution of nominal electron beam energies
in each pulse for the initial 100 and final 400 pulses in a burst,
corresponding to Fig. 3 (Suppl.).
The mean electron beam energy changed by about 30 MeV, or 0.22\%,
at the beginning of each burst.
The width of the nominal electron energy distribution
given by the last 80\% of the burst is 22 MeV FWHM, or 0.16\%.
Because the first harmonic photon energy emitted by the undulator
scales as the square of the electron beam energy,
the 0.22\% change in mean electron beam energy would produce a 0.44\% change in the
mean x-ray photon energy,
which is large compared to the 0.1\% bandwidth of the
first harmonic.
This produced a significant change in the x-ray pulse energy
at the photon energy selected by the monochromator;
the net effect on this experiment was that the average
available pulse energy 
was about $1 \times 10^{10}$ photons per pulse in our single pulse data,
and about $4 \times 10^{10}$ photons per pulse in our 500-pulse data. 

The wide variation in monochromatic x-ray pulse energy evident in Fig. 3 (Suppl.)
is also due to variations in the electron beam energy,
because the 0.16\% FWHM pulse-to-pulse fluctuations
in the nominal electron beam energy observed after the transient
will produce a 0.32\% fluctuation in the x-ray photon energy that is larger than the
0.1\% photon energy bandwidth of each x-ray pulse.
Figure 6 (Suppl.) shows the monochromatic x-ray pulse energy plotted against 
the nominal electron beam energy deviation from 13.426 GeV, for each pulse in a 500-pulse burst.
There is a strong correlation,
with a peak at about -6 MeV and a shoulder at about +6 MeV.
Figure 7 (Suppl.) shows the corresponding pink x-ray pulse energies plotted in the same way;
the lack of correlation indicates that it is not the number of x-ray photons, but the mean photon energy
in a pulse that varies to produce the peak in Fig. 6 (Suppl.).
Since the overall width of this peak is broader than the 6.7 MeV (0.05\%) FWHM expected
if all electrons in a pulse had the same energy,
it primarily reflects the distribution of electron energies within each pulse.
The distribution is inverted; e.g. the shoulder at higher nominal electron beam energy
arises from electrons with lower-than-nominal energy.

\begin{figure}
\includegraphics[width=3.0in]{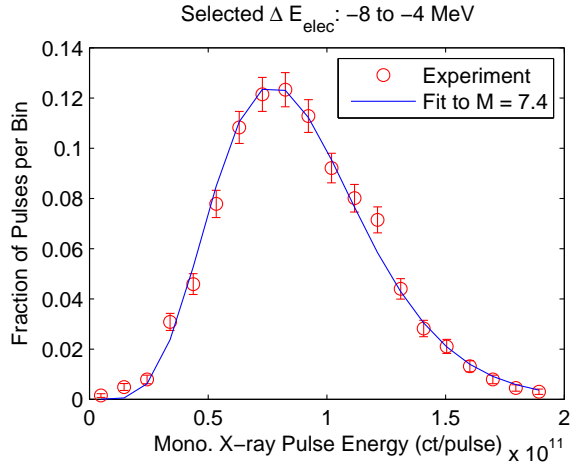}
\caption{(Supplemental) Histogram of monochromatic x-ray pulse energy of pulses selected to have
a nominal electron beam energy deviation in the range -8 to -4 MeV,
for 50 bursts.
Curve shows fit to Gamma distribution with mode number $M = 7.4$.}
\end{figure}

\begin{figure}
\includegraphics[width=3.0in]{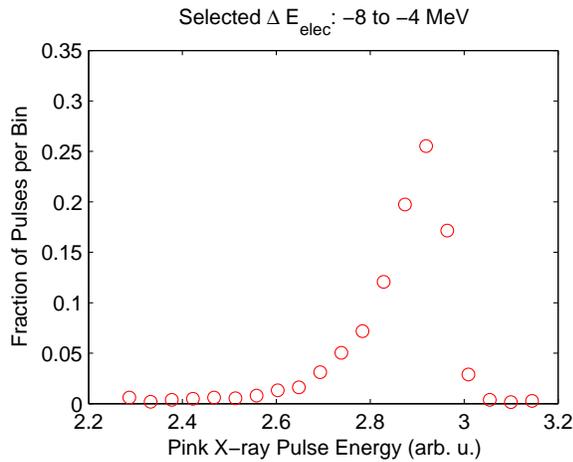}
\caption{(Supplemental) Histogram of pink (un-monochromated) x-ray pulse energies 
from the electron beam energy range indicated in Fig. 7 (Suppl.).}
\end{figure}

If we select pulses having a nominal electron beam energy within a narrow range,
the distribution of monochromatic x-ray pulse energies reflects the distribution expected from
the SASE FEL process, where the x-ray photon energy spectrum contains many ``spikes'' that vary
from pulse to pulse.
Figure 8 (Suppl.) shows such a distribution, for nominal electron beam energies between -8 and -4 MeV 
(covering the peak in Fig. 6 (Suppl.)).
The x-ray pulse energy distribution is no longer maximum at zero, as in Fig. 3 (Suppl.),
and can be fit by a Gamma distribution \cite{NB}
\begin{equation}
P_{\Gamma}(I) = \frac{M^M I^{M-1}}{\Gamma(M) \bar{I}^M} \exp(-M I / \bar{I})
\end{equation}
with a temporal mode number of $M = 7.4$.
This can be compared with the simple calculation of 18.5 modes 
for a Si (111) bandwidth ($1.4 \times 10^{-4}$ FWHM or 1.1 eV at 7.99 keV)
compared to a spike width of $\Delta E = h / \Delta \tau = 5.9 \times 10^{-2}$ eV using 
$\Delta \tau = 70$ fs from the electron pulse integral width.
While this is consistent with an effective x-ray pulse width that is about 40\%
of the 70 fs electron pulse width,
in rough agreement with other results \cite{GuttPRL2011, YoungNat2010, DuestererNJP2011},
the fit value of $M$ 
is a lower limit to the number of temporal modes in the x-ray beam 
(and thus the inferred x-ray pulse width is a lower limit) 
because of other potential contributions to the width of the x-ray pulse energy distribution.
The pink x-ray pulse energies corresponding to the same -8 to -4 MeV electron beam energy range 
are histogrammed in Figure 9 (Suppl.) 
and show a much more narrowly peaked distribution.

\section{Calculation of Maximum Speckle Contrast}

To perform XPCS on femtosecond time scales, which are much faster than the time resolution of imaging detectors,
a pulse split-and-delay technique has been proposed \cite{First_Experiments, GruebelNIM2007, GuttOE2009} 
in which the correlation time of the sample structure is determined 
from the change in contrast of the sum of two speckle patterns, 
as a function of the delay time between the patterns.
This is based on a property of speckle patterns --
the number of modes in a sum of equal photon density speckle patterns
is given by the sum of the number of modes in all the patterns if they are uncorrelated,
while it is equal to that of an individual pattern if they are fully correlated \cite{NB}.
Since the contrast factor $\beta$ is the inverse of the number of modes,
if we record the sum of two speckle patterns of contrast factor $M_0^{-1}$ 
from two equal-energy x-ray pulses separated by a time $\tau$,
the contrast factor of the sum will drop from $M_0^{-1}$ to $(2M_0)^{-1}$
as $\tau$ is varied from much less than to much more than the correlation time 
of the sample dynamics \cite{GruebelNIM2007}.

Since measuring speckle contrast is critical to ultrafast XPCS,
our experimental conditions were chosen to give high contrast
from the atomic-scale order in the samples.
This places conditions on the transverse and longitudinal coherence 
of the incident beam and the angular resolution of the detector \cite{PuseyReview, SuttonReview}.
Here we describe calculations of the maximum contrast factor
$\beta_{calc}$ we would expect to observe from a static sample
under our experimental conditions.
We can obtain an analytical formula for $\beta_{calc}$ in terms of
experimental parameters such as x-ray bandwidth, beam size and detector geometry
by considering the ratios between the experimental resolutions in reciprocal space 
and those required to fully resolve the speckle structure.

Since the transverse coherence length of the incident x-ray beam from LCLS is expected
to be the full size of the beam,
we simply assumed that the transverse coherence requirements are fully met
and do not contribute to decreasing $\beta_{calc}$ (increasing $M$).
This neglects any potential effects of aberrations in the beamline optics.
(Our observation of contrast factors approaching 90\% of the calculated $\beta_{calc}$ values
indicates that such effects are indeed small.)

The shape of a ``speckle'' (i.e. signal correlation volume) in reciprocal space
can be calculated from the shape of the scattering volume in real space.
This can be described by $t$, the effective length of the scattering volume along the beam direction,
and $s_v$ and $s_h$, the transverse beam sizes (FWHM) in and out of the scattering plane,
respectively.
The non-zero photon energy bandwidth of the incident beam $\Delta E / E$ 
gives an experimental resolution in the radial direction of
$\delta Q_r^{exp} = (\Delta E / E) Q$.
For transmission geometry,
the speckle correlation volume in reciprocal space is an ellipsoid 
with diameters $2\pi / t$, $2\pi / s_v$, and $2\pi / s_h$,
tilted at the Bragg angle
$\theta = \sin^{-1}(Q \lambda / 4 \pi)$
with respect to the radial direction,
so its radial width is
$\delta Q_r^{sp} = 2\pi / (s_v^2 \cos^2 \theta + t^2 \sin^2 \theta)^{1/2}$.
For symmetric reflection geometry,
this expression simplifies to
$\delta Q_r^{sp} = 2\pi / (t \sin \theta)$.
The factor by which $M$ is increased due to the non-zero radial resolution
can be expressed as
$M_{rad} = [ 1 + (\delta Q_r^{exp} / \delta Q_r^{sp} )^2 ]^{1/2}$,
which for transmission geometry gives
\begin{equation}
M_{rad}^{tr}
= \left [1 + \frac {Q^2 (\Delta E / E)^2 (s_v^2 \cos^2\theta + t^2 \sin^2\theta)}{4 \pi^2} \right ]^{1/2}
\end{equation}
and for symmetric reflection geometry gives
\begin{equation}
M_{rad}^{refl}
= \left [1 + \frac {Q^2 (\Delta E / E)^2 t^2 \sin^2\theta}{4 \pi^2} \right ]^{1/2} .
\end{equation}
The detector pixel size $p$ at a distance $L$ from the sample
gives a resolution of
$\delta Q_d^{exp} = 2\pi p / \lambda L$,
while the widths of the speckle ellipsoid in the detector plane are
$\delta Q_h^{sp} = 2\pi / s_h$ and
$\delta Q_v^{sp} = 2\pi M_{rad} / (s_v^2 \cos^2 2\theta + t^2 \sin^2 2\theta)^{1/2}$,
where here the effect of photon energy bandwidth on the speckle ellipsoid 
is included in the latter expression.
The factor by which $M$ is increased by the non-zero detector resolution
is given by
\begin{eqnarray}
M_{det} &=& \left [1 + \left (\frac{\delta Q_d^{exp}}{\delta Q_h^{sp}} \frac {\delta Q_d^{exp}}{\delta Q_v^{sp}} \right )^2 \right ]^{1/2} \\
&=& \left [1 + \frac{p^4s_h^2 (s_v^2 \cos^2 2\theta + t^2 \sin^2 2\theta)}{\lambda^4 L^4 M_{rad}^2} \right ]^{1/2}.\nonumber
\end{eqnarray}
The values of $M_{rad}$ and $M_{det}$ calculated for each of the samples and $L$ values,
and the net calculated contrast factor for a static sample, $\beta_{calc} = (M_{rad} M_{det})^{-1}$,
are given in Table I (Suppl.),
using the symmetric reflection geometry expression for the Ga and Ni$_2$Pd$_2$P samples,
and the transmission geometry expression for the B$_2$O$_3$ sample.
The calculated values of $\beta_{calc}$ are compared with the measured values of $\langle \beta \rangle$
in Table I and Fig. 3 of the main paper.

\begin{table}
\caption{(Supplemental) Calculated $M_{rad}$, $M_{det}$, and maximum contrast factor $\beta_{calc}$ for
Ga liquid, Ni$_2$Pd$_2$P glass, and B$_2$O$_3$ glass
at various detector distances $L$,
with $p = 20~\mu$m, $s_h = 1.6~\mu$m, $s_v = 1.7~\mu$m,
$\lambda = 1.552$~\AA, $\Delta E / E = 1.4 \times 10^{-4}$.
The effective sample length $t$ is taken to be half of the absorption length
for the reflection geometry samples (Ga and Ni$_2$Pd$_2$P),
and the sample thickness for B$_2$O$_3$ in transmission geometry.
}
\label{tab2}
\begin{ruledtabular}
\begin{tabular}{r|r|r|r|r|r|r}
Sample & $Q$ & $t$ & $L$ & $M_{rad}$ & $M_{det}$ & $\beta_{calc}$ \\
& (\AA$^{-1}$)& ($\mu$m)  & (cm)  &  & & \\
\hline
Ga & 2.60 & 14 & 37 & 2.79 & 1.17 & 0.307 \\
Ni$_2$Pd$_2$P & 2.95 & 4.1 & 18 & 1.40 & 2.05 & 0.349  \\
Ni$_2$Pd$_2$P & 2.95 & 4.1 & 35 & 1.40 & 1.11 & 0.645 \\
Ni$_2$Pd$_2$P & 2.95 & 4.1 & 142 & 1.40 & 1.00 & 0.713 \\
B$_2$O$_3$ & 1.62 & 18 & 37 & 1.75 & 1.28 & 0.446 \\
\end{tabular}
\end{ruledtabular}
\end{table}

\section{Extracting Contrast of Weak Speckle Patterns}

\subsection{Negative Binomial Distribution}

The signal distribution in a continuous speckle pattern with $M$ modes
is described by the Gamma distribution,
given in Eq. (1) above.
The normalized variance of this distribution
${\rm Var}_\Gamma / \bar{I}^2 = M^{-1}$
is equal to the contrast factor $\beta \equiv M^{-1}$ of the speckle pattern.
Thus the contrast factor can be extracted from intense speckle patterns
by evaluating the normalized variance in the signal distribution on the detector
\cite{GuttOE2009,GuttPRL2011,LudwigJSR2011}.
For weak speckle patterns,
the fluctuations due to photon counting statistics must be considered
\cite{NB, GuttOE2009,LivetAC2007}.
The probability distribution for the number of photons per pixel $k$ in a discrete speckle pattern
is described by
the negative binomial distribution \cite{NB},
which is the convolution of the Gamma and Poisson distributions,
\begin{equation}
P_{NB}(k) = \frac{\Gamma (k+M)}{\Gamma (M) \Gamma (k+1)} \left(1+\frac{M}{\bar{k}}\right)^{-k} \left(1+\frac{\bar{k}}{M}\right)^{-M}.
\end{equation}
The normalized variance of the negative binomial distribution is the sum of those
for the Gamma and the Poisson,
${\rm Var}_{NB} / \bar{k}^2 = M^{-1} + \bar{k}^{-1}$,
where $\bar{k}$ is the mean number of photons per pixel in the discrete speckle pattern.
For weak speckle patterns,
e.g. $\bar{k} < 1$,
the extra term $\bar{k}^{-1}$ from photon counting statistics
is larger than the contrast factor $M^{-1}$,
so that extracting the contrast factor from the normalized variance
can be inaccurate.
We instead can extract $\beta$ directly from the observed
$P(k)$ values using the negative binomial distribution.
In particular,
for small $\bar{k}$,
the probabilities for $k = 1$ and 2 photons per pixels
can be approximated by the leading terms in their series expansion in $\bar{k}$,
\begin{eqnarray}
P(1) &\approx& \bar{k} - (1 + \beta) \bar{k}^2 \\
P(2) &\approx& (1 + \beta) \bar{k}^2 / 2 - (1 + \beta)(1 + 2\beta) \bar{k}^3 / 2.
\end{eqnarray}
In this case the quantity $R \equiv 2 P(2) [1 - P(1)] / P(1)^2$
is a good estimator of $1 + \beta$.

The statistical error in using this estimate to extract $\beta$ 
from weak speckle patterns
is dominated by the error in the experimental determination of $P(2)$,
owing to the relatively small number of pixels with two photons.
For small $\bar{k}$, the number of two-photon events is given by 
$P(2) n_{pix} n_{patt} \approx (1 + \beta) \bar{k}^2 n_{pix} n_{patt} / 2$,
where $n_{pix}$ is the number of pixels in a pattern 
and $n_{patt}$ is the number of speckle patterns.
The relative error in this quantity due to counting statistics
is the inverse of its square root,
which gives the relative error in $R$ as 
$\sigma_R / R = [(1 + \beta) \bar{k}^2 n_{pix} n_{patt} / 2]^{-1/2}$
and the relative error in $\beta$ as
\begin{equation}
\frac{\sigma_{\beta}}{\beta} = \frac{1}{\beta \bar{k}} \left ( \frac{2(1 + \beta)}{n_{pix} n_{patt}} \right )^{1/2},
\end{equation}
which is the inverse of the signal-to-noise ratio for $\beta$ used in the main paper.

\subsection{Droplet Algorithm}

To obtain experimental values for $P(k)$, we first use a ``droplet'' algorithm
similar to that described previously \cite{Livet:2000p4334} to locate the positions 
of each photon detected by the CCD camera.
This is necessary because the signal for photons striking the detector
close to the boundary between pixels will have their signal spilt between pixels.
Each detector image was processed in the following way:

1) An averaged dark image was subtracted.

2) All pixels containing a signal less than a noise threshold of 5 ADU (analog to digital units) were set to zero.

3) Each connected region of pixels with non-zero signal was identified as a ``droplet.'' 
Connectivity was defined as adjacency within a row or column (not diagonally).

4) The signal within each droplet (ADU) was obtained by summing over the pixels in the droplet.

Figure 10 (Suppl.) shows a typical histogram of the signal in droplets.
There are sharp peaks at multiples of 764 ADU,
corresponding to integer numbers of 7.99 keV photons in each droplet.
This establishes the energy calibration of the detector and its electronics.
The width of the first peak (15 ADU FWHM) indicates an energy resolution
of 160 eV.
We also see ``escape peaks'' about 165 ADU below each of the lowest peaks,
corresponding to the loss of the energy of a Si K$\alpha$ fluorescence photon
(1.74 keV).

5) Each droplet was assigned an integer number of photons, 
based on its total signal and the location of the peaks in the histogram.
We set the boundary lines, shown in Fig. 7 (Suppl.), to be half way between the peaks.
The number of photons could thus be obtained by dividing the total signal
by 764 ADU and rounding to the nearest integer.
In particular droplets with signal below 372 ADU were taken to be noise
and assigned zero photons.
The choice of the signal values for the boundary lines for assignment of photon numbers
was varied by $\pm 200$ ADU and found to have minimal effect on the results.

6) To determine the location of the photons in each droplet 
we first separately consider the droplets falling into the following two simple cases:

- Case 1: If the droplet consists of only one pixel,
then the center of that pixel gives the position of all photons in the droplet.

- Case 2: If the droplet contains only one photon,
then the position of that photon is given by the $x$ and $y$ centroid  
positions of the signal in that droplet.

7) Each of the remaining multi-pixel, multi-photon droplets is analyzed
using the following procedure:

a) First a simple procedure is used to assign pixels to each photon,
to create initial guesses for a least-squares fitting routine.
The pixel with the highest signal is assigned a photon,
and 764 ADU is subtracted from that pixel.
This procedure is repeated until all photons have been assigned to pixels
in the droplet.  

b) Each droplet is considered in turn. 
The droplet is copied into a rectangular array that extends it
by zero-signal pixels, 
so that all sides are buffered by a border of zero-signal pixels.

c) A function containing $x$ and $y$ positions for each photon as parameters
is fit to the data in the array using a least-squares fitting routine. 
The function is built by assuming each photon produces uniform signal
in a square footprint the size of one detector pixel.
Its 764 ADU signal is split between four pixels based on the photon position
and the overlap of the area of this square with the underlying pixel array.

\begin{figure}
\includegraphics[width=3.0in]{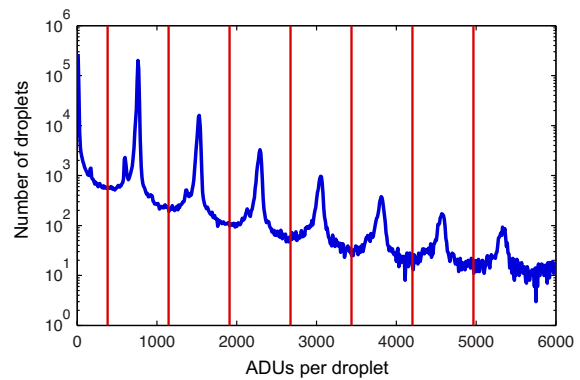}
\caption{(Supplemental) Histogram of analog-to-digital units (ADU)
per droplet, for all of the 309 images reported for liquid Ga.
Integer photon peaks occur at multiples of 764 ADU,
and Si K$\alpha$ escape peaks can be seen
below the primary peaks.
Vertical lines show boundaries used to separate droplets into
integer numbers of photons.
}
\end{figure}

d) The initial guess for photon positions in a droplet comes  
from the centers of the pixels assigned in step a), 
randomized by $\pm 0.5$ pixel in $x$ and $y$.

d) A weighted chi square representing the square of the deviation of the fit function from the data
is calculated for each step in the fitting process,
and the position parameters are varied until a minimum chi square is found.

e) The fitting procedure is run 4 times for each droplet, 
starting with a new randomization of the initial guess.
The positions from the fit with the minimum final chi square are chosen.

f) These photon positions relative to the droplet are reintegrated onto the full CCD array
based on the droplet position.
Photon positions from the special cases in step 6) are included.
The photon positions shown in Fig. 1(c) of the main paper were obtained in this way.

We found that this droplet algorithm worked efficiently 
for images with $\bar{k}$ up to about 0.2.
Above this value of $\bar{k}$,
the droplets began to grow much larger
and contain many photons
as the percolation limit was approached,
greatly slowing the fitting process.

\subsection{Calculation of $\langle \beta \rangle$}

Because both the incident pulse energy and effective photon energy bandwidth of each LCLS pulse
can be different \cite{GuttPRL2011},
we expect each speckle pattern to have different values of $\bar{k}$ and $\beta$.
Our strategy for extracting accurate contrast values using many weak speckle patterns
is thus to obtain relatively inaccurate $\beta$ values from each pattern
and to average them to obtain a more accurate overall $\langle \beta \rangle$.

To obtain $\beta$ for a given pattern from the photon positions,
we first chose a region of pixels with nearly constant $\bar{k}$
(e.g. along the top of the main amorphous scattering ring,
as shown in Fig. 1(a) of the main paper).
We used 300-, 600-, or 1100-pixel-wide regions
for the $L = 18, 35 ~{\rm to}~ 37, ~{\rm or}~ 142$~cm
sample-to-detector distances,
respectively.
The length of the region was 1200 pixels in all cases,
so that  the total numbers of pixels in the regions
were $n_{pix} = 3.6, 7.2, ~{\rm or}~ 13.2 \times 10^5$,
respectively.
These regions gave a variation of $\bar{k}$ of less than 5\%.

To calculate $P(k)$, each pixel within the chosen region
was assigned the number of photons that had positions
anywhere within its borders.
The experimental $P(k)$ for that image are given by 
the number of pixels with $k$ photons
divided by $n_{pix}$.
The experimental value of $\bar{k}$ for that image
was obtained from the total number of photons
divided by $n_{pix}$.

The method we used to obtain $\langle \beta \rangle$ from the $P(k)$
for a sequence of speckle patterns
was to calculate the quantity $R \equiv 2 P(2) [1 - P(1)] / P(1)^2$
for each pattern and perform a weighted average
to obtain a value of $\langle R \rangle$ for the set of patterns.
The weighting was based on the experimental uncertainties in $R$
which are dominated by the relatively small number of $k = 2$ pixels,
as described above.
Patterns with $\bar{k} > 0.2$ were not included,
because of errors introduced by the droplet algorithm.
Patterns with $\bar{k} < (2/n_{pix})^{1/2}$ were not included,
because below this limit the expected number of pixels with $k = 2$
becomes less than one for $\beta = 0$. 
The value of $\langle \beta \rangle$ was calculated from
$\langle R \rangle$
by numerically inverting the exact negative binomial 
expression for $2 P(2) [1 - P(1)] / P(1)^2$
using the average value $\langle \bar{k} \rangle$
for the set of patterns.
This corrects for the small deviation of $R$ from $1 + \beta$
at higher $\bar{k}$ values.
The RMS uncertainties in $\langle \beta \rangle$ reported in Table I
of the main paper were obtained by carrying the
experimental uncertainties through the weighted averaging procedure.

As can be seen from Fig. 2 in the main paper,
this method for obtaining $\langle \beta \rangle$
from the values of $P(2)$ and $P(1)$
gives good agreement with the observed values of $P(3)$ and $P(4)$
for single-pulse measurements.
Fitting of a single value of $\beta$ to all of the $P(k)$ values as a function of $\bar{k}$
gave an equivalent result for the single-pulse measurements.

However, we found that for the speckle patterns recorded using multiple LCLS pulses,
the observed values of $P(k)$ for $k > 2$ were typically
higher than predicted by the $\langle \beta \rangle$
extracted from $P(2)$ and $P(1)$.
This effect is consistent with the $P(k)$ predicted for the
summation of speckle from individual pulses with varying values of $\beta$
because of variations in the effective x-ray photon energy bandwidth.
Since the summation of patterns of different $\beta$
gives values of  $P(1)$ and $P(2)$ that agree with the average $\langle \beta \rangle$
to first order,
we chose to use the procedure outlined above to extract the contrast factors
for all of the data.

\section{Estimated Temperature Increase from X-ray Absorption}

Because the sub-picosecond x-ray pulse width is much smaller than the
time constant for thermal diffusion into the bulk from micron-scale volumes,
the energy deposited by x-ray absorption
can significantly heat the probed region of the sample
when focused LCLS beams are used.
This is a particular concern for XPCS measurements
where we wish to study the equilibrium dynamics of the sample
without perturbation by the x-ray probe.

For the Si (111) bandwidth monochromatic x-ray beam used here,
we find that typical unattenuated single pulses 
focused to a $1.6 \times 1.7~\mu$m FWHM spot
melted or vaporized the illuminated region of
Ga and Ni$_2$Pd$_2$P samples.
Figures 11-12 (Suppl.) show scanning electron microscope (SEM) images
of damage pits left in Ni$_2$Pd$_2$P glass by single LCLS pulses
of different pulse energy.
Major vaporization occurs for $I_0 = 2.4 \times 10^{10}$ photons
while only melting appears to have occurred for 
$3.3 \times 10^{9}$ photons.
Pulse energies of $2.1 \times 10^{9}$ photons and below left no mark
visible in SEM.

One can calculate the absorbed energy per atom in the illuminated volume as
$4 I_0 E_{phot}/(\rho_{at} \ell_{abs} \pi s_h s_v)$,
where $E_{phot}$ is the photon energy,
$\rho_{at}$ is the atomic density,
$\ell_{abs}$ is the incident x-ray absorption length,
and $s_h$ and $s_v$ are the transverse beam sizes.
The observed maximum pulse energy $I_0 = 2.1 \times 10^{9}$ 
that leaves no mark on the Ni$_2$Pd$_2$P sample
corresponds to 13 eV per illuminated atom absorbed energy.
Assuming a typical heat capacity of $3 k_B = 2.6 \times 10^{-4}$ eV/K per atom
one can estimate the adiabatic temperature rise that would occur 
if all of the absorbed energy remained in the illuminated volume.
This simple calculation gives
an adiabatic temperature rise of $5 \times 10^4$ K.

The thermal energy density needed to leave a damage mark
can be estimated by considering the time constant for viscous flow $\tau_{vf}$
relative to that for thermal diffusion $\tau_{th}$
calculated the measured properties of Ni$_2$Pd$_2$P.
Using a thermal diffusivity $D_{th} = 2.2 \times 10^{-6}$~m$^2$/s \cite{NPPdiffth},
the thermal time constant for dissipation of the heat from the x-ray pulse into the bulk
is $\tau_{th} = r^2/4D_{th} \approx 1~\mu$s
for a length scale $r$ of a few microns.
The time constant for viscous flow is \cite{Stephenson88}
$\tau_{vf} = 6 \eta (1-\nu) / E$,
where $\eta$ is the viscosity,
$\nu$ is Poisson's ratio,
and $E$ is Young's modulus,
all of which have been measured for Ni$_2$Pd$_2$P glass
 \cite{NPPvisc,NPPPois,NPPYoung}.
For a damage mark to occur, 
$\tau_{vf}$ must be much less than $\tau_{th}$,
which requires a temperature approaching 900~K
to get into the low-viscosity liquid 
$\eta < 10^4$~Pa-s  \cite{NPPvisc}.
Taking into account the doubling of the heat capacity at temperatures above
the glass transition at $\sim$550~K \cite{NPPheatcap},
we estimate that a damage mark will occur at deposited thermal energy densities
above about 0.26~eV per atom.
Comparing this to the observed damage threshold fluence of 13~eV per illuminated atom,
this indicates that the absorbed energy is thermalized in a volume 
about 50 times as large as the illuminated volume
for these micron-scale incident beams.

Because the x-ray absorption process involves a cascade of photoelectrons, 
secondary electrons, fluorescent photons, and optical phonons
that can propagate a significant distance before the absorbed energy
is transferred to a thermal population of lattice vibrations \cite{Cargill},
for small incident beams one indeed expects that the volume of material
that is initially heated to be larger than the illuminated volume.
The maximum temperature rise $\Delta T$ can be estimated as
\begin{equation}
\Delta T = \frac{I_0 E_{phot}}{3 k_B \rho_{at} \ell_{abs} \pi (s_h^2/4 + r_d^2)^{1/2}(s_v^2/4 + r_d^2)^{1/2}},
\end{equation}
where 
$r_d$ is the effective diffusion length of the non-thermal electron-phonon cascade.
For the beam size used here
($s_h = 1.6 ~\mu$m, $s_v = 1.7 ~\mu$m),
a cascade diffusion length of $r_d \approx 6 ~\mu$m
would agree with the observed factor of 50 larger volume
in which the absorbed energy is thermalized.
This is in rough agreement with the smallest melted regions observed,
e.g. Fig. 12 (Suppl.),
and with theoretical expectations,
since typical values for photoelectron propagation \cite{Cargill}
and electron excitation depths \cite{Potts}
are in the few-micron range.

\begin{figure}
\includegraphics[width=3.0in]{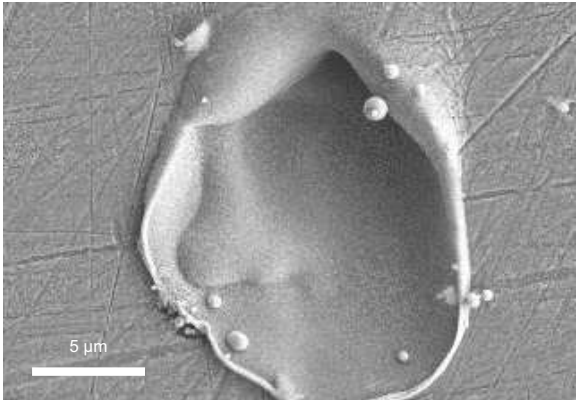}
\caption{(Supplemental) Scanning electron micrograph of 
damage pit 
left by single pulse of monochromatic 7.99 keV x-rays from LCLS
containing $2.4 \times 10^{10}$ photons (31 $\mu$J) 
focused to a $1.6 \times 1.7~\mu$m FWHM spot,
incident at 15 degrees onto the surface of Ni$_2$Pd$_2$P metallic glass
(scan 133 pulse 24).
The fluence of 1100 J/cm$^2$ and absorption length of $8.2~\mu$m
correspond to an average energy deposition of 150 eV per atom in the illuminated volume.
}
\end{figure}

\begin{figure}
\includegraphics[width=3.0in]{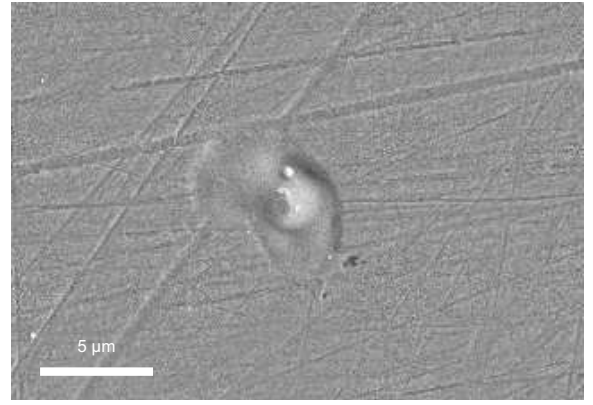}
\caption{(Supplemental) Scanning electron micrograph of 
damage pit 
left by single pulse of monochromatic 7.99 keV x-rays from LCLS
containing $3.3 \times 10^{9}$ photons (4.2 $\mu$J) 
focused to a $1.6 \times 1.7~\mu$m FWHM spot,
incident at 15 degrees onto the surface of Ni$_2$Pd$_2$P metallic glass
(scan 133 pulse 23).
The fluence of 155 J/cm$^2$ and absorption length of $8.2~\mu$m
correspond to an average energy deposition of 21 eV per atom in the illuminated volume.
}
\end{figure}

For the liquid Ga sample,
the vaporizations of the illuminated volume by intense single pulses
were observed to cause recoils
that changed the shape of the droplet, 
which then slowly returned to its original spherical shape.
We did not observe any visible change in the B$_2$O$_3$ fibers
even by the most intense pulses.

\section{Split-Pulse XPCS Feasibility}

One measure of the x-ray fluence that can be tolerated
can be obtained from our current measurements of the threshold of
pulse energy that reduced the speckle contrast in multiple-pulse patterns from
the glass samples.
We estimate that the threshold to permanently perturb the sample
structure corresponds to x-ray pulses of 
$I_0 = 1.6 \times 10^8$ or $1.4 \times 10^{10}$ photons of $E_{phot} = 7.99$ keV for
Ni$_2$Pd$_2$P or B$_2$O$_3$,
respectively.
These perturbation thresholds both correspond to 
an absorbed energy per illuminated atom of about 1 eV,
or a maximum temperature rise of about 80~K from Eq. (9),
owing to the different x-ray absorption lengths
of Ni$_2$Pd$_2$P and B$_2$O$_3$.
Pulses of this fluence caused no significant permanent rearrangement
of the atomic-scale structure e.g. through atomic diffusion.
However, when studying samples with more temperature-sensitive diffusion dynamics,
or more sensitive processes such as phonon dynamics,
an 80~K temperature rise may disturb the dynamics. 
Here we will estimate the minimum x-ray fluence and temperature rise needed
to obtain sufficient signal-to-noise to perform XPCS measurements.

The figure of merit for split-pulse XPCS measurements developed above,
$\beta / \sigma_{\beta}$,
is the same as that previously developed for standard ``sequential'' XPCS measurements
\cite{FalusJSR2006,LudwigJSR2011},
and the expression in terms of experimental parameters
$\beta / \sigma_{\beta} = \beta \bar{k} [n_{pix} n_{patt} / 2 (1 + \beta)]^{1/2}$
is very similar.
Both have the same linear dependence on $\beta \bar{k}$
and square root dependence on $n_{pix} n_{patt}$.
If sample perturbation by the x-ray beam is not an issue,
then a strategy to maximize the figure of merit
is to increase the mean photon density $\bar{k}$ at fixed $\beta$
by decreasing the focal spot size 
and increasing the angle $p/L$ subtended by each pixel,
even if the available solid angle is limited 
and the number of pixels $n_{pix}$ must thus decrease.

The mean photon density on the detector for a single pulse
can be extrapolated from our measured values using the expression
\begin{equation}
\bar{k} = \Sigma_v I_0 t \left ( \frac{p}{L} \right )^2,
\end{equation}
where $\Sigma_v$ is a sample-dependent cross section per unit volume.
By averaging the $\Sigma_v$ values obtained from our measured values of $\bar{k}$ per pulse
for each type of sample,
we obtain $\Sigma_v \approx 42$, 100, and 1.3 m$^{-1}$ rad$^{-2}$
for Ga, Ni$_2$Pd$_2$P and B$_2$O$_3$,
respectively.

In the limit of large detector pixel subtended angle $p/L$,
the expressions given above for $\beta_{calc}$ reduce to
$\beta = \lambda^2 (p/L)^{-2} s_h^{-1} (s_v^2 \cos^2 2\theta + t^2 \sin^2 2\theta)^{-1/2}$,
where $\beta$ is much smaller than unity.
Thus $\beta \bar{k}$ becomes independent of $p/L$, and is given by
\begin{equation}
\beta \bar{k} = \frac{\lambda^2 \Sigma_v I_0 t} {s_h (s_v^2 \cos^2 2\theta + t^2 \sin^2 2\theta)^{1/2}}.
\end{equation}
In this limit the signal to noise ratio $\beta / \sigma_{\beta} = \beta \bar{k} [n_{pix} n_{patt} / 2 (1 + \beta)]^{1/2}$
is given by
\begin{equation}
\frac{\beta}{\sigma_{\beta}}  = \frac{\lambda^2 \Sigma_v I_0 t} {s_h} \left ( \frac{n_{pix} n_{patt}} {2s_v^2 \cos^2 2\theta + 2t^2 \sin^2 2\theta} \right ) ^{1/2}.
\end{equation}

To estimate the signal-to-noise ratio for an optimized XPCS experiment at LCLS, 
we choose an incident x-ray pulse energy $I_0$
to produce a given temperature rise $\Delta T$
for a given beam size $s = s_v = s_h$ and sample
according to Eq. (9), giving
\begin{equation}
\frac{\beta}{\sigma_{\beta}}  = \frac{3 k_B \Delta T \lambda^2 \Sigma_v t \rho_{at} \ell_{abs} \pi (n_{pix} n_{patt})^{1/2} (s^2/4 + r_d^2)} 
{E_{phot} s (2s^2 \cos^2 2\theta + 2t^2 \sin^2 2\theta)^{1/2}}.
\end{equation}
Values of $Q \equiv 4\pi \sin \theta / \lambda$ and $t$ from Table I (Suppl.) are used, 
except $t = 200~\mu$m for B$_2$O$_3$.
We assume a detector capable of recording $n_{pix} = 10^6$ pixels at the full LCLS repetition rate of 120 Hz,
to give $n_{patt} = 4 \times 10^4$ in six minutes per delay time.
Figure 13 (Suppl.) shows the signal-to-noise ratio
as a function of beam size $s$
for a fixed $\Delta T = 10$~K,
for each of the three samples.
The solid curves correspond to the expected cascade diffusion length of $r_d \approx 6~\mu$m,
and show that the best signal to noise is obtained with small beam sizes.
This suggests that adequate signal-to-noise values (above the value of 5 shown by the dashed black line)
can be obtained for all samples in this regime.
At beam sizes smaller than $r_d$, 
the $I_0$ to give $\Delta T = 10$~K
from Eq.~(9)
becomes independent of beam size
and is given by 
$I_0 = 18, 0.5,~{\rm or}~0.2 \times 10^8$ photons per pulse
for B$_2$O$_3$, Ga, or Ni$_2$Pd$_2$P,
respectively.
Such pulse energies are expected to be available
even when using the narrow energy bandwidth of
$\Delta E / E = 10^{-5}$
from the pulse split-and-delay
\cite{RosOE2009,RosJSR2011}.

\begin{figure}
\includegraphics[width=3.0in]{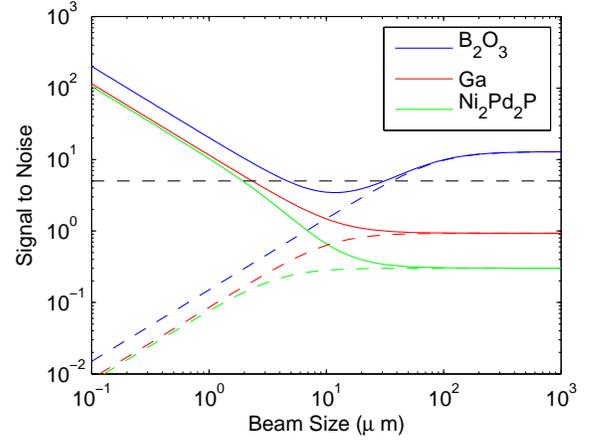}
\caption{(Supplemental) Estimated optimized signal-to-noise ratio as a function of beam size
for the three samples (Ga, Ni$_2$Pd$_2$P, and B$_2$O$_3$),
with pulse energy $I_0$ set to give a temperature rise of $\Delta T = 10$~K.
Solid curves are for a cascade diffusion length of $r_d = 6~\mu$m,
while dashed curves show the behavior for $r_d = 0$.
Black dashed line shows minimum required signal-to-noise ratio of 5.}
\end{figure}

\begin{figure}
\includegraphics[width=3.0in]{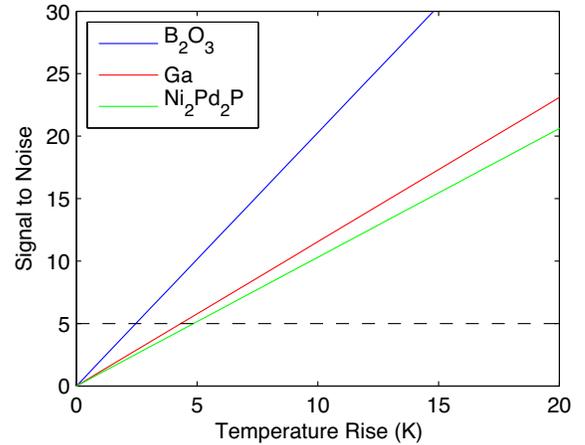}
\caption{(Supplemental) Estimated optimized signal-to-noise ratio 
as a function of temperature rise
for the three samples (Ga, Ni$_2$Pd$_2$P, and B$_2$O$_3$),
using a beam size $s = 1~\mu$m,
for $r_d = 6~\mu$m.
Black dashed line shows minimum required signal-to-noise ratio of 5.}
\end{figure}

The dashed curves on Fig. 13 (Suppl.) show the estimated behavior 
expected for the case of $r_d = 0$,
where all absorbed energy is thermalized within the illuminated volume.
In this limit, the best signal to noise is obtained for large beam sizes.
Adequate signal-to-noise values are still expected for samples
with low absorption such as B$_2$O$_3$.

Figure 14 (Suppl.) shows the signal-to-noise ratios as a function of temperature rise
for a fixed beam size of $s = 1~\mu$m,
for $r_d = 6~\mu$m.
The signal-to-noise ratio and temperature rise both scale linearly with $I_0$.
The estimated temperature rise needed to give a signal-to-noise ratio of 5
is $\Delta T = 3.3, 6.0,~{\rm or}~6.5$~K,
and the pulse energy is estimated to be $I_0 = 6, 0.3,~{\rm or}~0.13 \times 10^8$ photons per pulse,
for B$_2$O$_3$, Ga, or Ni$_2$Pd$_2$P,
respectively.
These temperature rises are significantly below the 80~K threshold
for permanent structural disturbance that we observed for the glass samples,
indicating that even more sensitive samples and processes can be studied.
By using beam sizes smaller than $1~\mu$m,
it may be possible to reduce $\Delta T$ even further.
However, ultrafast processes that disturb the sample structure 
on time scales shorter than that for thermalization of the absorbed energy
may become the limiting factor for very small beam sizes, as described below.

\section{Sample Disturbance During Single Pulses}

We observed that the average contrast factor $\langle \beta \rangle$ for single pulses 
was typically somewhat smaller than that calculated for a static sample, $\beta_{calc}$,
indicating that there could be some motion of the atoms
on the time scale of the x-ray pulse.
The ratio $\langle \beta \rangle / \beta_{calc}$ 
was smaller for Ni$_2$Pd$_2$P than for Ga,
and the average energy deposited per atom per pulse was larger,
consistent with motion driven by the energy of the x-ray pulse.
To evaluate the threshold above which
the atomic structure could be disturbed during the time scale of the pulse,
in Fig. 15 (Suppl.) we have plotted the ratio $\beta / \beta_{calc}$ vs. energy density deposited 
in the illuminated volume for each individual pulse
in the sequences for Ni$_2$Pd$_2$P and Ga taken with $L = 35$ cm.
While the uncertainties of the individual points are relatively large,
there is a clear trend toward lower contrast for higher deposited energy densities.
Energy densities above about 50 eV per atom reduce the single-pulse speckle contrast.
This illustrates the ability of speckle contrast to reflect atomic-scale, sub-picosecond dynamics,
and indicates that x-ray fluences corresponding to absorbed energy densities above 50 eV per atom
are likely to affect experiments in which single-pulse speckle patterns 
are analyzed to obtain atomic-scale structural information,
for x-ray pulses corresponding to 70 fs electron pulse widths.

\begin{figure}
\includegraphics[width=3.0in]{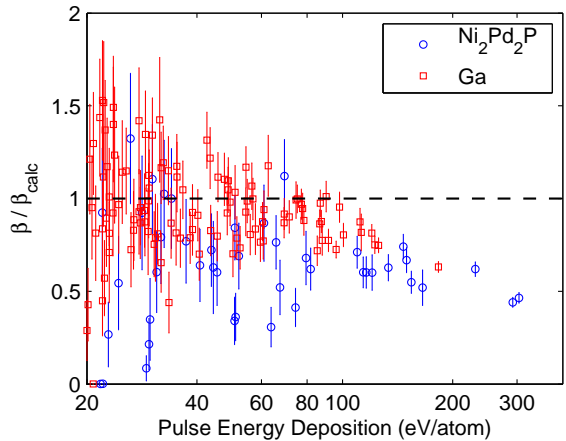}
\caption{(Supplemental) Ratio of extracted contrast factor $\beta$ to the calculated value $\beta_{calc}$
for single-pulse speckle patterns as a function of the x-ray pulse energy expressed as energy deposited per atom
in the illuminated volume. 
Data shown are from the sequences of single-pulse speckle patterns taken with L = 35 and 37 cm for Ga 
and Ni$_2$Pd$_2$P, respectively.}
\end{figure}

While the damage-mark threshold measurements described and modeled above
indicate that the energy absorbed in the illuminated volume can be thermalized
in a much larger volume for micron-scale beams,
thus reducing the sample disturbance,
ultrafast processes occurring prior to thermalization
can lead to a more restrictive intensity threshold for sample disturbance
for very small beams.
For example,
for the $I_0$ values used in Fig. 13 (Suppl.) which produce $\Delta T = 10$~K
according to Eq.~(9),
the threshold of 50 eV per illuminated atom
which we observe to disturb the structure during the x-ray pulse
will be exceeded for beam sizes smaller  than $s = 0.08~\mu$m.

\section{Acknowledgement}

Special thanks go 
to T. Hufnagel for the use of his metallic glass laboratory
and to A. Zholents for insightful comments.
X-ray experiments were carried out at LCLS at SLAC National Accelerator Laboratory, 
an Office of Science User Facility 
operated for the U.S. Department of Energy (DOE) Office of Science 
by Stanford University. 
Work at Argonne National Laboratory
supported by the DOE Office of Basic Energy Sciences 
under contract DE-AC02-06CH11357.  
Electron microscopy was accomplished at the Electron Microscopy Center for Materials Research
at Argonne, 
a DOE Office of Science User Facility.


\begin{thebibliography}{1}

\bibitem{NB}
J. W. Goodman,
\newblock {\it Speckle Phenomena in Optics : Theory and Applications} (Roberts \& Co., Englewood, Colo., 2007).

\bibitem{GuttPRL2011}
C. Gutt et al., Phys. Rev. Lett. {\bf 108}, 024801 (2012).

\bibitem{YoungNat2010}
L. Young et al., Nature {\bf 466}, 56 (2010).

\bibitem{DuestererNJP2011}
S. Duesterer et al., New Journal of Physics {\bf 13}, 093024 (2011).

\bibitem{First_Experiments}
G. B. Stephenson et al. in {\it LCLS: The First Experiments} (2000),
www-ssrl.slac.stanford.edu/lcls/papers/lcls\_experiments\_2.pdf.

\bibitem{GruebelNIM2007}
G. Gr\"{u}bel, G. B. Stephenson, C. Gutt, H. Sinn, and T. Tschentscher,
\newblock Nucl. Instrum. Meth. B {\bf 262}, 357 (2007).

\bibitem{GuttOE2009}
C. Gutt et al.,
\newblock Optics Express {\bf 17}, 55 (2009).

\bibitem{PuseyReview}
P.~N. Pusey,
\newblock Statistical properties of scattered radiation,
\newblock in {\em Photon Correlation Spectroscopy and Velocimetry}, edited by
  H.~Z. Cummings and E.~R. Pike, pages 45--141, Plenum, New York, 1974.

\bibitem{SuttonReview}
M.~Sutton,
\newblock Coherent x-ray diffraction,
\newblock in {\em Third-Generation Hard X-ray Synchrotron Radiation Sources:
  Source Properties, Optics, and Experimental Techniques}, edited by D.~M.
  Mills, John Wiley and Sons Inc., New York, 2002.
 
\bibitem{LudwigJSR2011}
K. Ludwig, J. Synchr. Rad. {\bf 19}, 66 (2012).

\bibitem{LivetAC2007} 
F. Livet, Acta Cryst. A {\bf 63}, 87 (2007).

\bibitem{Livet:2000p4334}
F.~Livet et~al.,
\newblock Nucl. Instrum. Meth. A {\bf 451}, 596 (2000).

\bibitem{FalusJSR2006}
P. Falus, L. Lurio, and S. Mochrie, J. Synchr. Rad. {\bf 13}, 253 (2006).

\bibitem{NPPdiffth} 
U. Harms, T.D. Shen, R.B. Schwarz,
Scripta Mater. {\bf 47}, 411 (2007).

\bibitem{Stephenson88} 
G. B. Stephenson, Acta Metall. {\bf 36}, 2663 (1988).

\bibitem{NPPvisc}
K. H. Tsang, S. K. Lee, and H. W. Kui,
 J. Appl. Phys. {\bf 70}, 4837 (1991).

\bibitem{NPPPois}
V. N. Novikov and A. P. Sokolov,
Phys. Rev. B {\bf 74}, 064203 (2006).

\bibitem{NPPYoung}
W. J. Wright, R. B. Schwarz, and W. D. Nix,
Mater. Sci. Eng. {\bf A319-321}, 229 (2001).

\bibitem{NPPheatcap}
G. Wilde, G. P. Gšrler, R. Willnecker, and G. Dietz,
Appl. Phys. Lett. {\bf 65}, 397 (1994)

\bibitem{Cargill}
A. Erbil, G. S. Cargill, R. Frahm, and R. F. Boehme,
Phys. Rev. B {\bf 37}, 2450 (1988).

\bibitem{Potts} P. J. Potts, {\em Handbook of Silicate Rock Analysis}, (Blackie, London, 1987), p. 336.

\bibitem{RosOE2009}
W. Roseker et al., Optics Lett. {\bf 34}, 1768 (2009).

\bibitem{RosJSR2011}
W. Roseker et al., J. Synchr. Rad. {\bf 18}, 481 (2011).

\end{thebibliography}
\end{document}